\documentstyle[aps,prl,epsf,floats]{revtex}
\newcommand{\la}[1]{\label{#1}}

\newcommand{\be}{\begin{equation}}
\newcommand{\ee}{\end{equation}}
\newcommand{\ba}{\begin{eqnarray}}
\newcommand{\ea}{\end{eqnarray}}
\newcommand{\bi}{\begin{itemize}}
\newcommand{\ei}{\end{itemize}}

\newcommand{\nr}[1]{(\ref{#1})}

\newcommand{\nn}{\nonumber \\}
\newcommand{\fr}[2]{{\frac{#1}{#2}}}
\newcommand{\msbar}{\overline{\mbox{\rm MS}}}

\newcommand{\<}{\left\langle}
\renewcommand{\>}{\right\rangle}

\newcommand{\bmu}{\bar{\mu}}

\newcommand{\eq}{Eq.~}

\newcommand{\fig}{Fig.~}

\def\lsi{\raise0.3ex\hbox{$<$\kern-0.75em\raise-1.1ex\hbox{$\sim$}}}
\def\gsi{\raise0.3ex\hbox{$>$\kern-0.75em\raise-1.1ex\hbox{$\sim$}}}
\newcommand{\lsim}{\mathop{\lsi}}
\newcommand{\gsim}{\mathop{\gsi}}

\begin{document}
\twocolumn[\hsize\textwidth\columnwidth\hsize\csname
@twocolumnfalse\endcsname

\title{A Strong Electroweak Phase Transition up to $m_H\sim 105$ GeV}
\author{M. Laine$^{\rm a,b}$ and K. Rummukainen$^{\rm c}$}
\address{$^{\rm a}$Theory Division, CERN, CH-1211 Geneva 23,
Switzerland}
\address{$^{\rm b}$Department of Physics,
P.O.Box 9, 00014 University of Helsinki, Finland}
\address{$^{\rm c}$NORDITA, Blegdamsvej 17,
DK-2100 Copenhagen \O, Denmark}
\date{April 7, 1998} 
\maketitle

\vspace*{-4.0cm}
\noindent
\hfill \mbox{CERN-TH/98-121, NORDITA-98/28P, hep-ph/9804255}
\vspace*{3.8cm}

\begin{abstract}\noindent
Non-perturbative lattice simulations have shown that there is no
electroweak phase transition in the Standard Model for the allowed
Higgs masses, $m_H\gsim 75$ GeV. In the Minimal Supersymmetric Standard
Model, in contrast, it has been proposed that the transition should 
exist and even be strong enough for baryogenesis up to $m_H\sim 105$ GeV, 
provided that the lightest stop mass is in the range 100\ldots160 GeV.
However, this prediction is based on perturbation theory, and suffers 
from a noticeable gauge parameter and renormalization scale dependence.
We have performed large-scale lattice Monte Carlo simulations
of the MSSM electroweak phase transition.
Extrapolating the results to the 
infinite volume and continuum limits, we find that the transition is 
in fact {\em stronger} than indicated by 2-loop perturbation theory. This 
guarantees that the perturbative Higgs mass bound $m_H\sim105$ GeV 
is a conservative one, allows slightly larger stop masses 
(up to $\sim$ 165 GeV), and provides a strong motivation for 
further studies of MSSM electroweak baryogenesis.
\end{abstract}

\vspace*{0.2cm}

\vskip1.5pc]

\noindent
It is known from studies of primordial nucleosynthesis
that there is a non-vanishing baryon to photon
density ratio in the Universe, $\eta\approx 10^{-10}$ 
(for recent reviews, see~\cite{bbn}). 
It is one of the main challenges of cosmology 
to understand how such an asymmetry could come about. 
Indeed, different scenarios for producing $\eta>0$ abound.

Among all the scenarios for baryogenesis, 
one is unique: the {\em last instance} 
in the history of the Universe that a baryon asymmetry could have
been generated, is the electroweak phase transition~\cite{krs}.
As such, this is also the scenario requiring
the {\em least assumptions} beyond established physics. 
In principle, even the Standard Model contains the necessary
ingredients for baryon number generation:
anomalous baryon number violation, CP-violation, 
and an electroweak phase transition providing for
a non-equilibrium environment (for a review, see~\cite{rs}).
Once an asymmetry has been generated, it must also be preserved, 
and this gives a strict constraint on how strongly of the
first order the transition must be~\cite{krs}.
In fact, the constraint on the strength of the phase transition
is the most rigorous of the constraints mentioned, since it concerns a 
thermodynamical {\em equilibrium} situation after the transition, 
and equilibrium physics is much better understood
than non-equilibrium physics.

However, it turns out that on a more quantitative level the Standard
Model is too restricted for baryogenesis.
The main reason is that the
strength of the electroweak phase transition depends on the 
Higgs mass, and for the allowed values $m_H\gsim 75$ GeV, 
there is no electroweak phase transition at
all~\cite{notransition,endpointmH}.
The existence of the baryon asymmetry 
alone thus requires physics beyond what is currently known.

The simplest extended scenarios that allow for baryon asymmetry
generation at the electroweak phase transition, have a Higgs sector
which differs from that in the Standard Model.  
A particularly appealing scenario 
is the electroweak phase transition in the
MSSM~[6--8]. 
Indeed, it has recently become clear that the electroweak phase
transition can then be much stronger than in the 
Standard Model, and strong enough for baryogenesis at least
for Higgs masses up to 80 GeV~[9--16]. 
For the lightest stop mass $m_{\tilde t_R}$ lighter than the top mass, one
can go even up to $\sim$ 100 GeV~\cite{bjls}: in the
most recent analysis~\cite{cqw2}, the allowed window was estimated 
at $m_H \sim 75\ldots105$ GeV, $m_{\tilde t_R} \sim
100\ldots160$ GeV. In this regime, the transition could
even proceed in two stages~\cite{bjls}, via an exotic
intermediate colour breaking minimum. This Higgs and stop mass
window is interesting from an experimental point
of view, as well, as the whole range will be covered
at LEP and the Tevatron~\cite{cqw2}.

Unfortunately, the statement concerning
the strength of the electroweak phase transition 
in this regime is subject to large uncertainties. 
The first indication in this direction is that the 2-loop
corrections to the Higgs field effective potential are large 
and strengthen the transition considerably~\cite{e}. 
A further sign is that the gauge parameter and, in particular,
the renormalization scale dependence of the 2-loop potential, 
which are formally of the 3-loop order, are numerically quite
significant~\cite{bjls}. Hence a non-perturbative analysis
is needed.

The purpose of this paper is to study 
the MSSM electroweak phase transition
with lattice Monte Carlo simulations, and to extrapolate
the results to the infinite
volume and continuum limits. Since the MSSM at finite temperature
is a multiscale system with widely different scales
from $\sim \pi T$ to $\sim g_W^2 T$, and since there
are chiral fermions, the only way to do the simulations
in practice is to use an effective 3d theory~\cite{4dsim}.
This approach consists of a perturbative dimensional reduction into a 3d
theory with considerably fewer degrees of freedom than in
the original theory~[21--23], 
and of lattice simulations in the effective theory. The analytical dimensional
reduction step has been performed for 
the MSSM in~[12--14,17]. 
Lattice simulations in dimensionally reduced 3d theories
have been previously used to determine
the properties of the electroweak phase transition in the
Standard Model in great detail~[24--30]. 

In the regime considered, the right-handed stop field $U$
plays an important role in addition to the Higgs field.
The effective 3d Lagrangian describing the electroweak
phase transition in the MSSM is therefore an SU(3)$\times$SU(2)
gauge theory with two scalar fields~\cite{ml,bjls}:
\ba
{\cal L}_{\rm cont}^{\rm 3d} & = &
\fr14 F^a_{ij}F^a_{ij}+\fr14 G^A_{ij}G^A_{ij}+ \gamma H^\dagger H U^\dagger U
\nn 
& + & (D_i^w H)^\dagger(D_i^w H)+
(D_i^s U)^\dagger(D_i^s U)+m_{H3}^2 H^\dagger H \nn 
& + &  m_{U3}^2U^\dagger U+ \lambda_{H} (H^\dagger H)^2+
\lambda_{U} (U^\dagger U)^2. \la{Uthe}
\ea
Here $D_i^w$ and $D_i^s$ are the 
SU(2) and SU(3) covariant derivatives, and
$H$ is the combination of the Higgs doublets which is ``light'' at the 
phase transition point.
The U(1) subgroup of the Standard Model induces only small
perturbative contributions~\cite{su2u1}, and can be neglected.

The complexity of the original 4d Lagrangian is hidden in 
\eq\nr{Uthe} in the expressions of the parameters of the 3d theory.
A dimensional reduction computation leading to actual
expressions for these parameters has been 
made in~\cite{bjls} for a particularly simple case. 
Let us stress here that the reduction
is a purely perturbative computation and is free of 
infrared problems. The relative error has been estimated 
in~\cite{ml,bjls}, and should be $\lsim 10\%$.

It is prohibitively time-consuming to study the full parameter space
of \eq\nr{Uthe} with Monte Carlo simulations.  Thus, we only consider
a special parameter choice: 
we take a large left-handed squark mass
parameter $m_Q\sim$1 TeV, vanishing squark mixing parameters, and a
heavy CP-odd Higgs particle ($m_A \gsim 300$ GeV).  We fix
$\tan\beta=3$, corresponding to $m_H\sim 95$ GeV.  We then study the
3d theory in \eq\nr{Uthe}, parametrized by the temperature $T$ and the
right-handed stop mass parameter $\tilde m_U$ ($\tilde m_U$ determines
the zero temperature right-handed stop mass through $m_{\tilde
t_R}\approx ({m_{\rm top}^2-\tilde m_U^2})^{1/2}$).  The actual
expressions used for the dimensional reduction are given
in~\cite{lr2}.

The philosophy is now that we determine the non-perturbative
results for the continuum theory in \eq\nr{Uthe} through lattice
simulations, and compare them with 3d perturbation
theory, employing the same 3d parameters. To be more precise, we 
compare with 2-loop 3d perturbation theory in the Landau gauge $\xi=0$
and for the $\msbar$ scale parameter $\bmu=T$, values which
have been used in~\cite{cqw2}, as well. This allows one to find out whether 
there are any non-perturbative effects in the system. Once this has 
been done, one can go back to a more complicated situation and study it
perturbatively, adding to the perturbative results the non-perturbative
effects found here. 
As the reduction step is 
purely perturbative, the non-perturbative effects found with 
the 3d approach apply also to the effective potential 
computed in 4d~\cite{e,ce,cqw2}.

To perform lattice simulations, 
we discretize the theory in \eq\nr{Uthe} 
with standard methods (see~\cite{lr2}). 
The lattice parameters are expressed
in terms of the lattice spacing $a$ and the 
continuum parameters through 2-loop relations~\cite{ref:hpert3}
which become exact in the continuum limit.

Well controlled {\em infinite volume}
and {\em continuum} limits are essential in order to
obtain reliable results.  Thus, for each point in the parameter
space, we always perform simulations
with several lattice volumes and extrapolate to the infinite
volume. We use the lattice spacings obtained through
\be
\beta_S\equiv \fr{6}{g_{S3}^2 a}=12,20,
\ee
where $g_{S3}^2=g_S^2T\sim T$ is the 3d SU(3) gauge
coupling and $a$ is the lattice spacing. The fact that
we use just two values of $\beta_S$, 
only allows a linear extrapolation to the continuum limit 
$\beta_S=\infty$.  However, it
is understood analytically that the dominant corrections are
linear~\cite{moore2}, and moreover, linear extrapolations work
extremely well for the case of the Standard
Model~\cite{nonpert,su2u1}. 

All in all, we have performed 42 different Monte Carlo runs:
combinations of lattice sizes and parameters.  The total cpu-time was
$\sim$7.5 node-years on a Cray T3E\@.

The physical quantities we discuss here are the critical 
temperature $T_c$, the scalar field expectation values, 
and the latent heat. Quantities such as the latent heat 
enter, for instance, the estimates for the nucleation 
and reheating temperatures (see, e.g.,~\cite{hks}), which are needed
to decide whether the scalar field expectation values relevant
for cosmology should be taken at $T_c$ or some lower temperature.

\begin{figure}[t]
\centerline{\epsfxsize=7.6cm\epsfbox{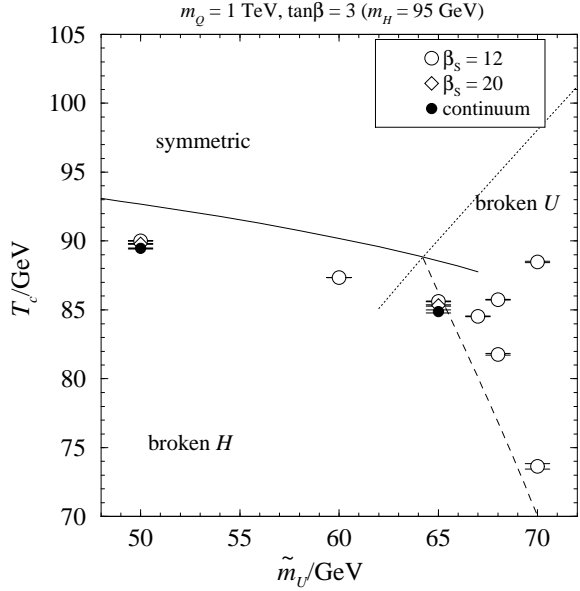}}
\caption[a]{The phase diagram and the critical temperatures.
The continuous lines are from the 2-loop perturbative 
effective potential in the Landau gauge.  Open symbols correspond to
infinite volume extrapolations, and filled symbols to 
continuum limit extrapolations.}\la{fig:tc}
\end{figure}

\begin{figure}[t]
\centerline{\epsfxsize=8cm\epsfbox{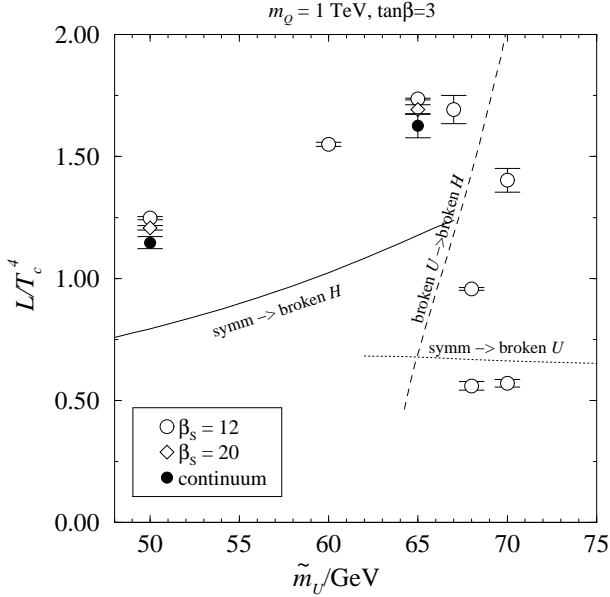}}
\caption[a]{The latent heat.}
\label{fig:latent}
\end{figure}

\begin{figure}[t]
\centerline{\epsfxsize=8cm\epsfbox{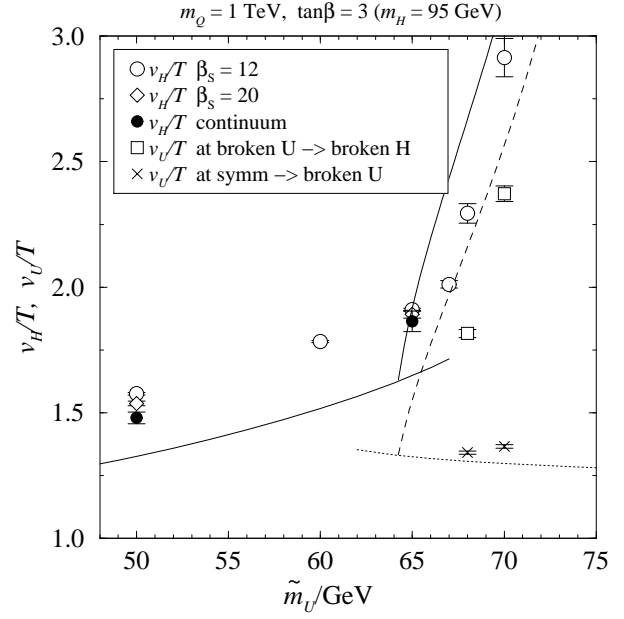}}
\caption[a]{The scalar field expectation values in
the broken phases at $T_c$.}\la{fig:vH}
\end{figure}

\vspace{2mm}
\noindent
{\em 1. The phase diagram and the critical temperatures.~~}\\
The general phase structure of the theory is expected to 
be the following~\cite{bjls}. The system  
has a first order transition at 
$T_c \sim 100$ GeV for $\tilde m_U \lsim 65$ GeV.
This transition is strong even though $m_H$ 
is large, due to the stop
loops. As $\tilde m_U$ becomes larger 
($m_{\tilde t_R}$ smaller), the transition gets even
stronger, and then at some point one may get a two-stage 
transition. The existence of a two-stage transition
depends on the parameters of the theory, and for large squark 
mixing parameters the two-stage region is not reached~\cite{cqw2}.

Our numerical
results are shown in \fig\ref{fig:tc}. 
It is seen that the phase diagram is 
qualitatively the same as in perturbation theory, although
the critical temperatures and the triple point have been displaced
by a few GeV\@. 
We have data at $\beta_S=20$ only at $\tilde m_U = 50$, 65\,GeV\@,
and the continuum extrapolation is possible only at these points.
Nevertheless, we expect similar (small) effects at the other points.
As of now, we have no clear theoretical explanation for
the discrepancy between the lattice results and perturbation theory:
the reason might be, e.g., a three-loop perturbative 
effect, or a genuine non-perturbative contribution.

\vspace{2mm}
\noindent
{\em 2. Latent heat.~~} \\
The main result of this paper is shown in \fig\ref{fig:latent},
which shows the latent heat. It is the most important 
gauge-invariant physical characterization of the strength 
of a first order transition. We observe that the non-perturbative
transition to the standard electroweak minimum at $\tilde m_U \lsim 67$ GeV
is significantly (up to 45\%) {\em stronger} than the 
perturbative transition.
In the regime $\tilde m_U\gsim 67$ GeV where there is a two-stage
transition, a comparison with perturbation theory is more difficult
as the whole pattern is shifted to the right, but the qualitative
behaviour is the same.

\vspace{2mm}
\noindent
{\em 3. Scalar field expectation values.~~}\\
The Higgs field vacuum expectation value $v_H$
is the object by which one usually characterises whether the phase
transition is strong enough for baryogenesis~\cite{krs,rs}, the 
requirement being $v_H/T\gsim 1$. As such $v_H$ is, however, a gauge 
dependent quantity. If one computes it in the Landau gauge ($v_H^L$), 
as is usual, then in terms of gauge-invariant operators the same 
expression would be non-local. On the other hand, there is a 
simple local gauge-invariant quantity closely related 
to $v_H$, namely $H^\dagger H\sim v_H^2/2$.  The problem
with $H^\dagger H$ is that being a composite operator, 
it is a scale dependent quantity in, say, the $\msbar$ scheme.
We hence define on the lattice
\be
\frac{v_H}{T}  \equiv  
\left( 2 
\< \frac{H^\dagger H_{\overline{\mbox{\scriptsize\rm MS}}}
 (g_{S3}^2)}{T} \> \right)^{1/2}, \la{vHT}
\ee
which is a natural gauge-invariant generalization of $v_H^L/T$, 
and can be measured in simulations.
Note that with respect to 4d units, there is a trivial 
rescaling by $T$ in the $H^\dagger H$ appearing in \eq\nr{vHT}.

The numerical results for $v_H/T,v_U/T$ are shown 
in \fig\ref{fig:vH}. Again, we observe a value larger
than in perturbation theory in the regime $\tilde m_U\lsim 67$ GeV.
Moreover,  in qualitative accordance with 
perturbation theory, there is 
a rapid increase in $v_H/T_c$ in the regime of the two-stage transition, 
$\tilde m_U\gsim 67$ GeV. The relative non-perturbative
strengthening effect is smaller than for the latent heat, 
which is easy to understand since
$L\propto \Delta (H^\dagger H) \sim \Delta v_H^2$~\cite{nonpert}, 
implying $\delta L/L \sim 2 \delta v_H/v_H$.

In conclusion, at least 
for the parameter values studied ($m_H\sim 95$ GeV, 
$m_{\tilde t_R}\sim 150\ldots 160$ GeV), the electroweak
phase transition is significantly stronger than indicated by 
2-loop perturbation theory. This implies
that the previous perturbative Higgs and stop mass bounds 
for electroweak baryogenesis
are conservative estimates. In particular, 
the electroweak phase transition could be
strong enough for baryogenesis for {\em all allowed 
Higgs masses} in this regime ($m_H\lsim 105$ GeV)~\cite{cqw2}. 
Due to the non-perturbative strengthening effect seen, 
the stop mass could be slightly larger than the perturbative value, 
up to $m_{\tilde t_R}\sim 165$ GeV. 
For the smallest stop masses, on the other hand, 
there is the possibility
of a two-stage transition, in which the Higgs field gets
an extremely large vacuum expectation value.

These results provide a strong 
motivation for precise studies of the non-equilibrium 
CP-violating real time dynamics and baryon number generation
at the MSSM electroweak phase transition. 

{\bf Acknowledgements.}
The simulations were made with a Cray T3E at the 
Center for Scientific Computing, Finland. 
We acknowledge useful discussions with 
K. Kajantie, G.D. Moore, M. Shaposhnikov 
and C. Wagner. This work was partly supported by the
TMR network {\em Finite Temperature Phase Transitions
in Particle Physics}, 
EU contract no.\ FMRX-CT97-0122.

\end{document}